# Preliminary study on the modal decomposition of Hermite-Gaussian beams via deep learning

Yi An[†], Tianyue Hou[†], Jun Li, Liangjin Huang**, Jinyong Leng, Lijia Yang and Pu Zhou*

*College of Advanced Interdisciplinary Studies, National University of Defense Technology, Changsha 410073, China*
*\* zhoupu203@163.com*
*\** hlj203@nudt.edu.cn*
*[†] These authors contributed equally to this work*

**Abstract:** The Hermite-Gaussian (HG) modes make up a complete and orthonormal basis, which have been extensively used to describe optical fields. Here, we demonstrate, for the first time to our knowledge, deep learning-based modal decomposition (MD) of HG beams. This method offers a fast, economic and robust way to acquire both the power content and phase information through a single-shot beam intensity image, which will be beneficial for the beam shaping, beam quality assessment, studies of resonator perturbations, and other further research on the HG beams.

## 1. Introduction

Hermite-Gaussian (HG) [1] and Laguerre–Gaussian (LG) beams[2], usually generated from the solid-state lasers, are gaining increasing interest in both fundamental and applied research fields. They are related to orbital angular momentum (OAM) beams[3], optical communications[4] and quantum information processing[5]. Recent years, much attention has been given to generate[6], detect[7], control[8], sort[9] or demultiplex[10] the modes of HG and LG beams. Besides these fruitful and beneficial results, modal decomposition (MD) is another essential and interesting topic, which can reveal the modal weights and phase information of the beams. With this information, much physical property of the field can be inferred, for example, wavefront[11], OAM density[12], and the beam quality[13]. Since the HG beams can be converted to LG beams[8] and two kinds of beams are similar to some extent, we focus on the modal decomposition of HG beams in this paper.

To decompose the HG beams, various methods have been proposed such as coherence measurements[14, 15], correlation filter techniques[16] and intensity recordings[17, 18]. Among these approaches, coherence measurements allow for the modal weights to be determined[14, 15], but the optical field cannot be reconstructed totoally due to the lack of phase information. Correlation filter techniques utilize computer-generated holograms (CGHs) to perform MD, which can obtain not only the modal weights but also the modal phases. Moreover, the high processing frequency of this technique allows for real-time modal analysis[16]. However, generating an appropriate CGH is costly and one piece of CGH is only suitable for the corresponding type of resonator. Although the phase-modulation spatial light modulator can serve as a CGH to overcome these drawbacks[19], it is relatively expensive. Intensity recording methods might be the most economical way to decompose HG beams since the ordinary Charge Coupled Device (CCD) camera can complete the recording tasks. In Ref. [17], it requires a handful of transverse intensity profiles sampled at various positions so that more time is required. Besides, this method [17] can only determine the modal weights so that the intensity profiles are impossible to be reconstructed to verify the accuracy, which hinders its practical application. In Ref. [18], both the near-field and far-field are utilized to calculate the modal weights and phase, but it is also relatively time-consuming due to its iterative calculation process.

Convolutional neural network (CNN)[20] is one kind of deep learning technique, which can learn the complex mapping between different domains through a training process. In this paper, we combine CNN with the MD of HG beams for the first time, which only requires one beam intensity to obtain both the modal weights and phases information. Moreover, with a trained CNN, the non-iterative MD can be completed in tens of milliseconds. This approach paves a way for simple, economical and fast real-time modal analysis of HG beams.

## 2. Method

### 2.1 The basics of modal decomposition

HG beams can be generated from laser resonators with rectangular geometry and the field distribution of HG modes can be expressed as[21]

$$HG_{mn}(x,y) = \frac{1}{w_0}\sqrt{\frac{2}{2^{m+n}\pi m!n!}} H_m(\frac{\sqrt{2}}{w_0}x) H_n(\frac{\sqrt{2}}{w_0}y) \exp(-\frac{x^2+y^2}{w_0^2}) \qquad (1)$$

where $w_0$ denotes waist radius of the fundamental mode $HG_{00}$ and $H_l$ is Hermite polynomial of order $l$. The intensity profiles of the former sixth HG modes are illustrated in Fig. 1.

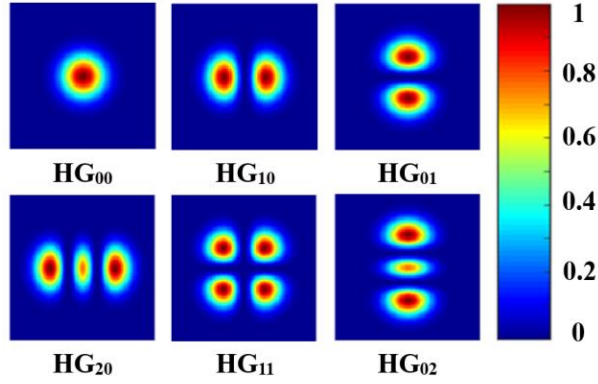

Fig. 1. The intensity profiles of the former sixth HG modes.

HG modes make up a complete and orthonormal basis so that an arbitrary transverse field $U$ can be expanded into a superposition of these modes[16]

$$U(x,y) = \sum_{m=0}^{\infty}\sum_{n=0}^{\infty} c_{mn} HG_{mn}(x,y), \qquad c_{mn} = \rho_{mn} e^{i\theta_{mn}} \qquad (2)$$

$$\sum_{m=0}^{\infty}\sum_{n=0}^{\infty} \rho_{mn}^2 = 1, \qquad \theta_{mn} \in [-\pi, \pi] \qquad (3)$$

where $\rho_{mn}^2$ denotes the modal weights while $\theta_{mn}$ is the modal phases. Since only the relative phases between the fundamental mode and higher-order modes make sense, we set $\theta_{00}$ as zero and then $\theta_{mn}$ represent the relative phases. The modal weights and the phases form the mode coefficient, which is expected to be determined through an arbitrary HG beam intensity.

*2.2 Principle of the scheme*

The principle of our scheme is shown in Fig. 2. The intensity profile of the HG beam emitting from the laser source is fed into the trained CNN to give the mode coefficient directly. The intensity image can be recorded by a CCD camera while CNN can be stored in a common computer, which are hidden in the figure. Since only a CCD and a common computer are required to perform MD, our scheme is very simple and economic.

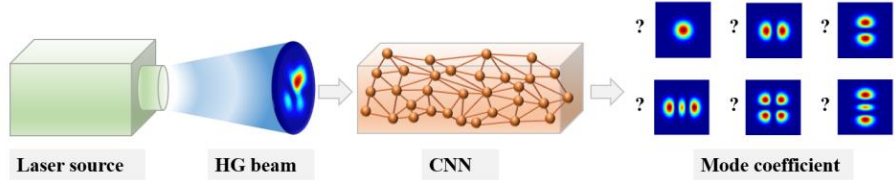

Fig. 2. The principle of our scheme.

The most important work of our scheme is to train the CNN previously. Training the CNN requires large amounts of samples, including near-field beam intensity images and their corresponding labels. The sample images can be easily acquired from Eq. (1)-(3) by varying $\rho^2$ and $\theta$ with

$$I(x,y) = |U(x,y)|^2 \qquad (4)$$

As for the labels, they are vectors ranging from [0,1] contain the modal weights and the linearly scaled cosine values of modal phases. The cosine function is adopted to avoid the conjugated field cases, which means one beam intensity corresponds to two field distribution whose sign of phases is totally opposite and more details about the case can be found in our previous work[22].

With the prepared samples, we modified VGG-16 model[23], which is a mature and widely used CNN architecture, to perform the training process. The first layer's filter size of VGG model is changed from 3×3×3 to 3×3×1 since our input is a single intensity pattern image while the last layer's Softmax function is replaced by Sigmoid for our regression problem. In the training process, the input sample images are passed through the layers of the CNN and regressed into an output vector. The mean-square error (MSE) between the output and the label vector is defined as the loss of CNN, then the parameters of the CNN are updated iteratively using back-propagated gradients based on the MSE loss.

When the network gets convergence after several training epochs, it can be utilized for predicting the mode coefficient through a non-iterative process. Taking a gray image of HG beams as input, modal weights can be predicted directly from the output vector of the trained CNN. For the modal phases, all the possible combinations based on the estimated cosine values are collected as candidates, and the final predicted phases can be determined by searching the maximum of the correlation between these candidates and the input intensity image[22].

**3. Results and discussion**

We take the former sixth HG modes as an example to verify the principle of our scheme. In every training epoch, 50000 beam intensity images with a resolution of 128×128 are randomly generated to train the CNN. The training is on a desktop computer with an Intel Core i7-8700 CPU and GTX 1080 GPU and the learning rate is set as 0.01. After 10 training epochs with about 2-hour cost time in total, the CNN gets convergence. Then we use 1000 simulated testing beam samples those are not included in the training datasets to evaluate the performance of the trained CNN.

Figure 3 shows three typical results of the testing samples, in which the intensity or phase map reconstructed by the MD results are illustrated to be compared with the actual ones. It can be found that the reconstructed intensity of patterns (A)-(C) are of great similarity to the input actual ones, indicating high accuracy. As for the reconstructed phase map, (A) and (B) agree well with their actual one while (C) is totally opposite to the actual one, which comes from the conjugated cases mentioned in Section 2.2 and it cannot be avoided since only one intensity image is utilized for decomposition[18]. The far-field beam intensity can be observed to distinguish these conjugated pairs.

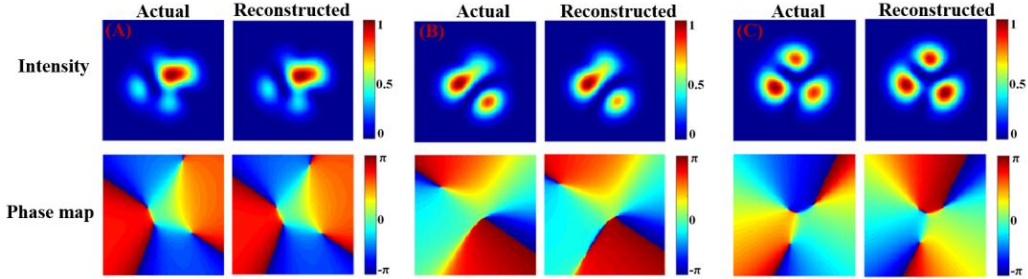

Fig. 3. Three typical results of the testing samples. The beam intensity and phase map are reconstructed based on the predicted mode coefficient, the actual ones are also shown for comparison.

We also compare the predicted and actual mode coefficients of these three typical samples, as shown in Fig. 4. The predicted modal weights of every mode are very close to the actual ones for all the three samples, so as the modal phases of pattern A and B, while the absolute value of predicted phases of pattern C is very close to the actual one, although the sign of them is opposite.

To assess the whole accuracy of 1000 samples, we define weights error $\Delta\rho^2$, phase error $\Delta\theta$ and prediction error (PE) of $i^{th}$ sample as Eq. (5) –(7) respectively. The averaged value for these defined errors are reported in Table 1 while the averaged consuming time for one image is only about 23 ms, which is very fast.

$$\Delta\rho_j^2[i] = \left|\rho_{j,a}^2[i] - \rho_{j,p}^2[i]\right| \quad j = 00,\ 10,\ 01,\ 20,\ 11,\ 02 \quad (5)$$

$$\Delta\theta_j[i] = \left\|\theta_{j,a}[i]\right| - \left|\theta_{j,p}[i]\right\|/2\pi \quad j = 10,\ 01,\ 20,\ 11,\ 02 \quad (6)$$

$$PE[i] = \frac{1}{11}\sqrt{\sum_j (\Delta\rho_j^2[i])^2 + \sum_j (\Delta\theta_j[i])^2} \qquad j = \begin{cases} 00, 10, 01, 20, 11, 02, & \text{for } \Delta\rho_j^2 \\ 10, 01, 20, 11, 02, & \text{for } \Delta\theta_j \end{cases} \qquad (7)$$

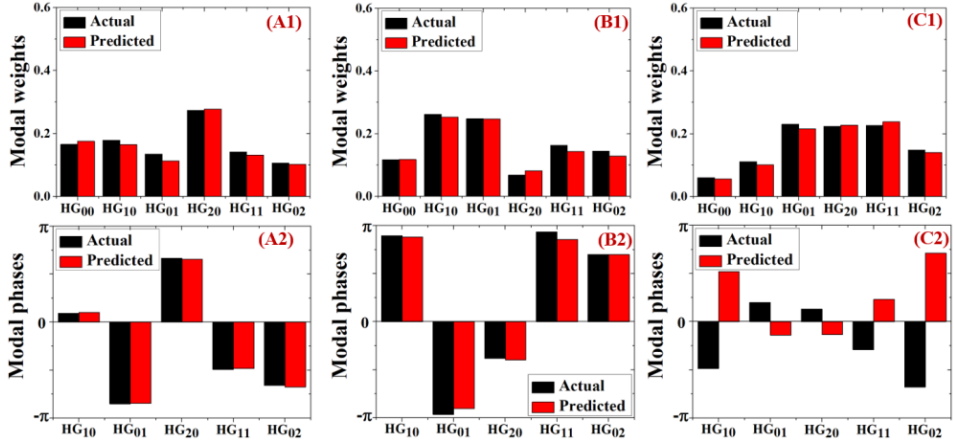

Fig. 4. Comparison of the predicted and actual modal weights or phases of three typical samples.

Table 1 Averaged value for defined error of 1000 testing samples.

| Error item | $\Delta\rho_{00}^2$ | $\Delta\rho_{10}^2$ | $\Delta\rho_{01}^2$ | $\Delta\rho_{20}^2$ | $\Delta\rho_{11}^2$ | $\Delta\rho_{02}^2$ |
|---|---|---|---|---|---|---|
| Averaged value | 0.0355 | 0.0355 | 0.0360 | 0.0185 | 0.0324 | 0.0164 |
| Error item | PE | $\Delta\theta_{10}$ | $\Delta\theta_{01}$ | $\Delta\theta_{20}$ | $\Delta\theta_{11}$ | $\Delta\theta_{02}$ |
| Averaged value | 0.0133 | 0.0392 | 00377 | 0.0439 | 0.0441 | 0.0442 |

Noted that the defined PE considers both the predicted weights error and phases error, we adopt this error to evaluate the accuracy of 1000 testing samples. The PE of these testing samples is plotted with red spots in Fig. 5 and the blue curve in the figure represents the averaged value. The reconstructed and actual intensity images of the $200^{th}$, $400^{th}$, $600^{th}$, $800^{th}$ and $1000^{th}$ samples are also illustrated in the insets of the figure, corresponding to (a)-(e) respectively. The small fluctuation of the PE and the high agreement between the reconstructed and actual intensity shows the trained CNN can achieve great accuracy.

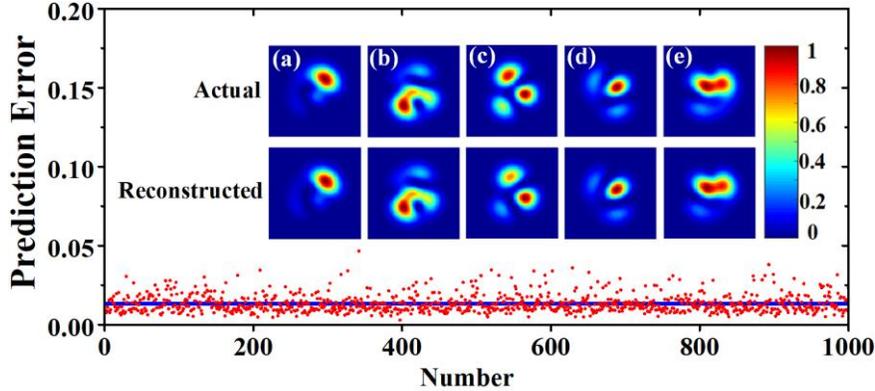

Fig. 5. The prediction error of 1000 testing samples. The red spots are the separate value and the blue curve represents the averaged value. The actual and reconstructed intensity profiles (a)-(e) correspond to the $200^{th}$, $400^{th}$, $600^{th}$, $800^{th}$ and $1000^{th}$ sample.

The robustness of CNN is investigated by adding noise to the input intensity image. The noise was added to the intensity value of each sample as a random number with a uniform distribution in the interval $(0, \varepsilon I_0)$, where $I_0$ is the maximum value of the intensity profile and $\varepsilon$ is a positive quantity. As shown in Fig. 6, when we increase $\varepsilon$ from 0 to 0.2, the averaged PE increases from 0.013 to 0.037 slowly, which means the scheme's accuracy will be influenced by the noise but can still achieve accurate MD results under moderate levels of noise. The insets of Fig. 6 illustrate a typical group of input intensity images under different noise levels and their corresponding reconstructed images. The $\varepsilon$ value of the beam profiles (a)-(f) are set as 0, 0.04, 0.08, 0.12, 0.16 and 0.2 respectively, among which intensity image (a) is the ground truth. It can be found that although the trained CNN can weaken the influence of noise, the reconstructed intensity image is not in superb agreement with the ground truth, especially under heavy noise. To further enhance the MD accuracy for noisy images, the noisy samples can be included in the training samples.

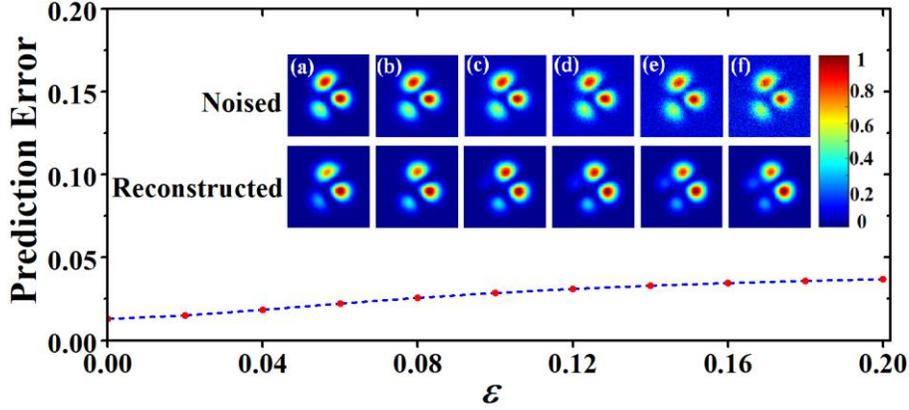

Fig. 6. The averaged prediction error of 1000 testing samples under different noise levels. The $\varepsilon$ value of the beam profiles (a)-(f) are set as 0, 0.04, 0.08, 0.12, 0.16 and 0.2 respectively.

In this paper, we investigate the deep learning-based MD of HG beams containing six basic modes through simulation, and the CNN is trained by large amounts of simulated samples. According to our previous work[22], the simulated samples are effective for the training of the prediction CNN, which is able to accurately analyze the experimental data. Therefore, when the scheme is utilized for real HG beams, the CNN can also be trained based on the simulated dataset, provided that the experimental conditions are taken into consideration, e.g., waist radius of the fundamental mode. Noted that we only discuss the 6-mode case of HG beams, if the scheme is applied to a higher number of modes, more training samples are needed to achieve the comparative accuracy due to the complexity of the intensity images when more modes are involved. Further, due to the similarity and the close relationship between the LG modes and HG modes, our approach can also be extended to LG beams.

Compared with existing MD methods for HG beams [14-18], our scheme is fast, simple, economical and robust, which is promising for real-time beam analysis, and will be beneficial to beam quality evaluation, studies of resonator perturbations and other further research on the intrinsic properties of HG beams

## 4. Conclusion

In this paper, we have proposed the modal decomposition of HG beams via deep learning for the first time, which is demonstrated through 6-mode simulated beam intensity images. Our approach only requires a single-shot beam image, so that greatly reduce the operation efforts and consuming time. With a trained CNN, both the modal weights and phases can be determined in about 23 ms from a prepared intensity image. When more modes are investigated, more training samples are required to train CNN for accurate results because the patterns will be more complex.